# Challenges and Opportunities Associated with Technology-Driven Biomechanical Simulations.

## Authors

1. Zartasha Mustansar[1], *Department of Computational Engineering, School of Interdisciplinary Engineering and Sciences, National University of Sciences and Technology* (NUST)*, Islamabad, Pakistan, zmustansar@sines.nust.edu.pk*
2. Haider Ali, *Department of Computational Engineering, School of Interdisciplinary Engineering and Sciences, National University of Sciences and Technology, Islamabad, Pakistan,* hdrali89@gmail.com
3. Lee Margetts, *Department of Mechanincal and Aerospace Engineering, University of Manchester UK. Lee.margetts@manchester.ac.uk*
4. Saad Ahmad Khan, *Department of Computational Engineering, School of Interdisciplinary Engineering and Sciences, National University of Sciences and Technology, Islamabad, Pakistan,* sahmed.mscse16@rcms.nust.edu.pk
5. Salma Sherbaz, *Department of Computational Engineering, School of Interdisciplinary Engineering and Sciences, National University of Sciences and Technology, Islamabad, Pakistan, salmasherbaz@sines.nust.edu.pk*
6. Rehan Zafar Paracha, *Department of Computational Engineering, School of Interdisciplinary Engineering and Sciences, National University of Sciences and Technology, Islamabad, Pakistan, rehan@sines.nust.edu.pk*

*[*1]Contact Author: Dr Zartasha Mustansar; email: zmustansar@sines.nust.edu.pk*

*Regular Research Paper*

## Abstract

This paper presents the principal challenges and opportunities associated with computational biomechanics research. The underlying cognitive control involved in the process of human motion is inherently complex, dynamic, multidimensional, and highly non-linear. The dynamics produced by the internal and external forces and the body's ability to react to them is biomechanics. Complex and non-rigid bodies, needs a lot of computing power and systems to execute however, in the absence of adequate resources, one may rely on new technology, machine learning tools and model order reduction approaches. It is also believed that machine learning approaches can enable us to embrace this complexity, if we could use three arms of ML i.e. predictive modeling, classification, and dimensionality reduction. Biomechanics, since it deals with motion and mobility come with a huge set of data over time. Using computational (Computer Solvers), Numerical approaches (MOR) and technological advances (Wearable sensors), can let us develop computationally inexpensive frameworks for biomechanics focused studies dealing with a huge amount of data. A lot of misunderstanding arises because of extensive data, standardization of the tools to process this, database for the material property definitions, validation and verification of biomechanical models and analytical tools to model various phenomena using computational and modeling techniques. Study of biomechanics through computational simulations can improve the prevention and treatment of diseases, predict the injury to reduce the risk and hence can strengthen pivotal sectors like sports and lifestyle. This is why we choose to present all those challenges and problems associated with biomechanical simulation with complex geometries fail so as to help improve, analysis, performance and design for better lifestyle.

**Keywords:** Computational biomechanics, challenges, technology, computing and simulations.

# 1. Introduction

The human systems are a composition of largely intricate assembly of bones, tissues and ligaments. Its structure provides a remarkable combination of rigidity, stability, and flexibility. Rigidity provides an essential vertical bony axis, stability provides strong scaffolding for cavities and extremities, and flexibility permits complex movements. Nature operates on the same principle to maintain a stable ecosystem which has always been an inspiration for the engineers to create optimized, reliable, efficient, human approved designs for movement and gait. This was always in perspective because, the continuum of the universe is typically very non-linear pattern and solving it on a brink of computers bring up a lot of challenges. Human bones, especially long ones are like beams to a huge body architecture. If we have a fragile or unstable core including our hips, it impacts our movement patterns, leading to an imbalance and injury pruned structure. At the same time, taking into account typically everything anatomically realistic is fairly impossible due to huge surge of dealing with computational powers. This is a big challenge to biomechanics' of today's era. The forces we usually encounter, during a normal activity of life include, ground reaction force, gravitational force, muscle forces and or forces along with body momentum as well. In early $20^{th}$ century, bionics appeared as a field related to technology inspired by biology. The transfer of technology from life forms to engineering gave birth to biomechanics. However, it has been very difficult to study through experimentation because of variety and uniqueness in biomechanical systems. In late $20^{th}$ century, Computer Aided Engineering (CAE) has come up as a key player in design decisions through its ability to minimize noise and statistical uncertainties. Inclusion of CAE in biomechanics started a new stream biomechanical simulations which does not require any experimentation. It became the most used technique to study the biomechanics related to systems. Its advantages include, not limited to, better understanding of biomechanical systems, fast analysis, and prediction of associated injury even under hypothetical conditions, modeling and extrapolating the biomechanics for a cohort, and making design decisions. However, advantages come with many challenges. Where researchers are finding it very useful, they are also encountering challenges. [1]–[4]. This paper therefore, addresses the challenges faced by the researchers while studying biomechanics using simulations. Few of grand challenges are discussed as under.

# 2. Challenges Associated with Biomechanics Model Building

Anatomical features of life forms make the biomechanical models complex which is a basic requirement for the computational biomechanical studies [5], [6]. Therefore, regular CAD modeling software cannot be used in modeling of biological systems. Different special featured software along with imaging modalities are used for modeling. However, it becomes a challenge as the biomechanical system does not exist independently [7]. Also, with the population the variation in models is large making the researchers to compromise with the error induced while using averaging techniques. Conversion of data from motion capture modalities also becomes a challenge when making it workable for the existing software. For collaboration of communities, standards must be adopted and compatibilities among different medical modalities and simulation software need to be assured. It has been presented in meetings of Grand Challenges on the Modeling and Simulation (M&S) as Big Simulation. It describes the issues related to big input data and large sets of coupled simulation models [1]. The data gathered from the imaging modalities usually contains details up to parts per billions hence making it very large to handle. Therefore, hi-Tech systems are required to analyze the data [8].

Recently, a study by [17] discusses motion capture systems. The way these tech-driven systems retrieve kinetic (and kinematics) data with less expert knowledge and without expensive equipment is mind-blowing. These motion capture systems are meant to increase the availability of motion analysis to a wider range of people, i.e. enabling systems for pervasive healthcare systems. Therefore, one of the biggest challenge to the biomechanics community is to deliver, right data in a smart way.

## 3. Finite Element Modeling on the move

Finite element modeling is playing a great role to control the computing aspects of biomechanical simulations and decrease the solution time according to the technology needs. But since FEM is an approximation method [reference] therefore currently two major issues of finite element method are accuracy and precision in relation to validation. This is sometimes due to a lack of computing power and/or computational cost. As the number of finite elements in a model increases, and the complexity of the physics included in the simulation increases, the computational load increases [9].

Computational requirements increase very quickly with increasing element counts. Model validation involves assessing the degree to which the finite element model (and the results that it outputs) represents the real system being modelled. In simple words, the analyst must assess whether the model correctly represents the geometry, loading conditions, material properties (and correct constitutive model), boundary restraints and interface conditions of the real structure. The accuracy of a finite element model is determined by performing convergence tests. A convergence test is carried out by analyzing essentially the same problem a number of times, but with increasing numbers of finite elements [9].

## 4. Challenges Associated with Material Property Definition

Every biological component in a system shows different mechanical properties even in a cohort of same species. Finding exact material properties e.g. elastic modulus is a difficult process. There are number of factors which affect the method of determining value of elastic modulus that may in turn affect elastic properties of bone. For example, a healthy will function much better than a damaged bone whereas a damaged bone is likely to cause a functional disability. So if condition of specimen changes, its functionality also changes. According to wolf's law, the mechanical environment also defines the properties of every bone in a living body. Over the year, mechanical properties have been quantified for few components through experimental studies on cadavers. However, properties for most of the biomaterials are still unknown. Therefore, assumptions are made, and closely related engineering materials are used instead. The task of assigning material properties of the elements is another major aspect of model generation in biomechanical simulations.

Various studies have been conducted on assigning the material properties to the model and directly addressed that in what way difference in material properties affects the performance of the model. Then these studies computed the results that effects on model behavior depend on the modification in degree of isotropy and regularity in the model. It has been observed that variation in material properties cause the several configurations of deformation in the models, although to gather such models is very time consuming but they seems to be vital for building a precise model [10]–[12]. Where to model and simulate the actual results need of accurate properties is inevitable. Therefore, it has become a challenge to address in the modeling and simulation community. As an example, see table 1- it is the data from one of our Emu biomechanics work, where we wanted to model the motion mechanics, using an MRI scan of an Emu bone and build a Finite Element-based model. Over time, we see change in the weight of the bone and hence the calculations. Therefore, one of the challenges is to see how varying bone weight can change the properties of the bone affecting simulation process data, hence outcomes.

| Date | Method of measurement | Weight in grams |
|---|---|---|
| **Day 1** | **Electrical balance** | **23.5885g** |
| **Day 2** | **Electrical balance** | **23.5577g** |
| **Day 3** | **Electrical balance** | **23.54562g** |
| **Day 4** | **Electrical balance** | **23.5422g** |

Table 1: Material Property Variation over time, while bone drying procedure

# 5. Machine Learning (ML) in Biomechanics – Opportunities and challenges

With the advent of latest technology and computing algorithm, complex non-linear procedures which are very dynamic in nature, are becoming relatively easier. ML algorithms have a wide range of applications and due to their fast computation times, their non-linearity as well as their adaptability to real and complex problems, its perhaps the finest choice by seasoned scientists. Because of the similar reason, therefore ML has already found some applications in biomechanics. Using ML, data acquisition in biomechanics can be handled more efficiently. Processes like optimization of inertial sensors is also possible due to ML ([16]. For the science of motion mechanics as well as sports, 3D kinematics and vertical ground reaction forces could be predicted from 2D videos and vice versa. But of course, these advantages do not come without challenges and drawbacks: We need a large amount of training data with a high changeability that describes the real problem as good as possible. Not only this, we also need to choose the right model to train, configure the parameters optimally and design the training process in such a way that there is no under- or overfitting. In practice, there are no exact guidelines that you can follow, so you have to use the trial-and-error method until you develop a feeling for the best settings. But all of this can be mastered, and then machine learning definitely has the potential to significantly increase the objectivity of decision-making in biomechanics and solve complex problems quickly in the future. Currently, what we follow as standard method of implementing ML ops to Biomechanics led projects, is shown in Fig (1).

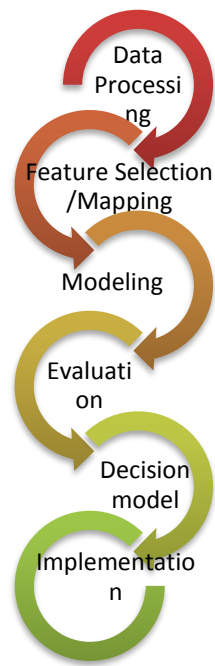

**Figure 1: Data challenge lifecycle of ML in Biomechanics**

Another example, for instance is exploiting the power and versatility of artificial (deep) neural networks as universal function approximators. Since motion analysis is a very trivial process of collecting angular moments, joint moments and kinematics data, an example of external knee moments calculation [18], [19], including running mechanics and using a minimal set of IMUs tells that this great challenge can be dealt with reasonably using approaches such as to augment a measured inertial sensor dataset with simulated data and then to demonstrate how to efficiently estimate sagittal plane angles, joint moments, and ground reaction forces.

## 6.     Challenges Associated with Imaging Data

This challenge is of prime importance. A lot of data which coming from Imaging modalities, do not reflect actual anatomical details due to various reasons including no access to the high end imaging modalities equipment's. When the subject is moving, the calculation points are not very clear as required. Given the variety of imaging modalities to-date and many more in the pipe line it leaves the clinical investigator in a dilemma regarding which imaging outcome measure to select for a particular study. Another similar challenge is that the sequences and most importantly the hardware are continually changing.  One of the examples is the use of MRI to evaluate the integrity of region of interests in clinical data especially with pathologies. Given in the figure is one of the examples from segmentation of human spine from our biomedical engineering lab explaining the content on our segmentation strategy. Due to low resolution image, segmentation was an overwhelming process. Overcoming the barrier of low resolution images has been a complex challenge always to the biomedical researchers. Low image quality yield to poor segmentation hence poor analysis.

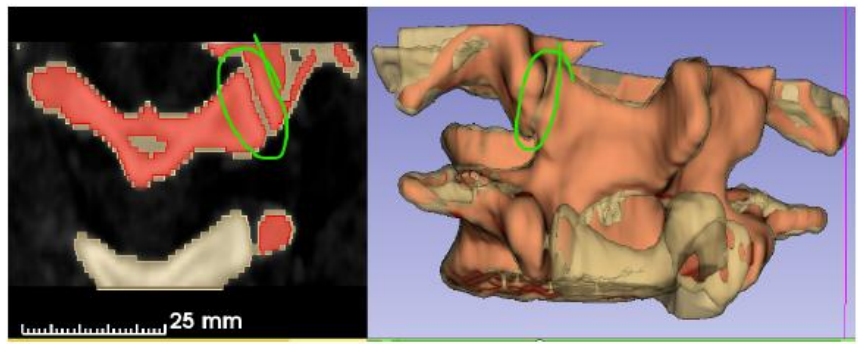

Figure 2: Image segmentation of human spine vertebra

Another important reason why we see quality of imaging in biomedical engineering is to verify if even there is a possibility of creating a Finite Element Model from this dataset or not. To validate this, usually an engineer does a smart test. An example of the same study can be cited here (fig 3) from biomechanics of EMU study as shown (Fig 4) – this image as you see is an example of a very low resolution image with faulty image information.

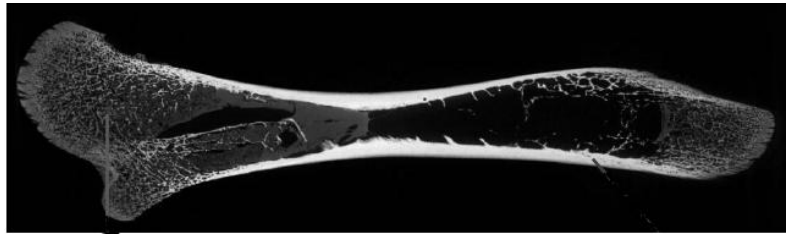

Figure 3: MRI scan of full femur of an Emu.

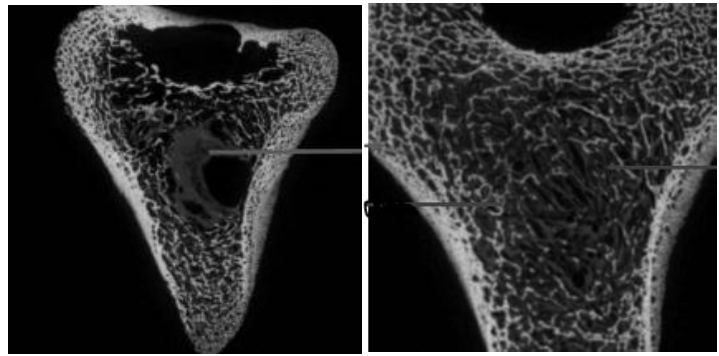

Figure 4: Cut through a raw slice of MRI scan from upper condyle of an Emu Femur.

To address this, a small cube was cut in our lab at Manchester, to see the dimension of the smallest possible feature. The diameter of the thinnest structure was found to be 2 pixels wide however for a good quality FE model, almost 5-6 pixel thick trabeculae was required (Industrial Finite element method standards). Moreover, there were pixel elements can be seen having absolutely no connectivity with each other which can later, break the high performance simulation due to faulty elemental connectivity problems, if this phenomenon was overlooked. These challenges, if addressed well before time, can save a lot of simulation time and can bring in a sophisticated technology in practice.

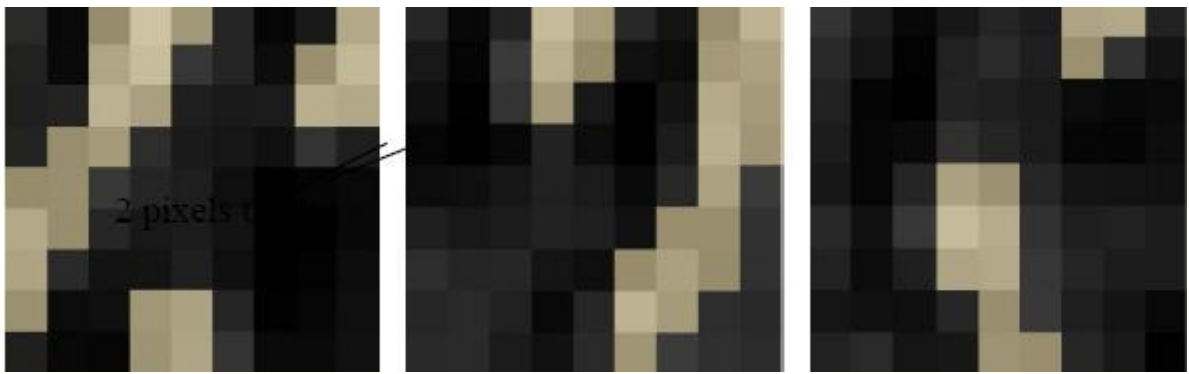

Figure 4: Pixelated Emu femur's cube data

## 7. Challenges Associated with Verification and Validation

Verification and validation (V&V) include the assessment of accuracy and validity of the computer models and computational predictions. Model verification deals with implementation of the model and its numerical accuracy where as the model validation deals with how well the model represents the real data [13]. Again, this is not easy step for a successful simulation study. In biomechanics, usually and extensively used models are established CAE methods i.e., finite element analysis of deformations in bones and tissue, multibody simulation techniques for musculoskeletal simulations and finite volume analysis for blood flow and respiration. However, with every new technology, an added complexity is catered in these models. Therefore, the methods and models which are accepted and reliable for the engineering problems may not fit completely suitable in the biomechanical simulations. There exists a lack of confidence in determining the inputs of the model which makes it difficult to obtain the accurate experimental data. Furthermore, the fear of losing the vitality of results from the model is what makes researchers more delusional. There exists but very limited literature on the validation within biomechanics based studies. Qualitative validation is possible for the available trends of certain components in given conditions where the literature lacks the quantitative analysis due to very complex, non-linear data-driven models. Therefore, indirect validation is mostly in practice as instances for direct validation is very rare [14][15]. For indirect validation similar conditions and materials in engineering are identified and studied. But still, it is challenging to look for the materials on par with the biological materials. Almost every possible hypothetical situation can be studied through biomechanical simulations. However, it will be difficult to reproduce the same situation in experiment and in clinical studies especially when predicting the injuries under certain given conditions.

## 8. Recommendations to the Scientific Community

After careful literature review and analysis, it is very important to understand to develop a pipeline for all these challenges to be dealt practically and with comparatively a short solution time. Challenges associated with data, is a real game changer. Data, if processed closely accurate, can bring out solutions which can improve the healthcare procedures and help design better equipment's. Apart from this, challenges associated with computing integrated with motion analysis, backed up by expensive motion capture systems, can be optimally integrated with Machine Learning approaches, inertial wearable sensors to efficiently devise health equipment's. Moreover, the ML approaches, can deliver us the best simulation decision models to prefer one simulation using a certain set of parameter over the other to reduce simulation time and hence computational cost as well. The integration of machine learning with biomechanics not only simplifies the assessment of several interdependent parameters but also provides the opportunity for automated and unbiased analysis.